\documentclass{article}

\usepackage[square,sort,comma,numbers]{natbib}
\usepackage{braket}
\usepackage{layout}
\usepackage{amsmath,amssymb,amsbsy}
\usepackage{listings}
\usepackage{todonotes}

\begin{document}

\lstset{basicstyle=\footnotesize,breaklines=true}

\title{Three Quantum Programming Language Parser Implementations for the Web}

\author{Marcus Edwards
\affil{Electrical and Computer Engineering,
University of British Columbia, Vancouver, BC, CA}}

\author{Marcus Edwards}

\maketitle

\begin{abstract}

IBM has developed a quantum assembly (QASM) language particular to gate model quantum computing since 2017 \cite{2017arXiv170703429C}. Version 3.0 which adds timing, pulse control, and gate modifiers is currently undergoing finalization in 2023 \cite{2021arXiv210414722C}. In a similar vein, Pakin of Los Alamos National Laboratory published a quantum macro assembler (QMASM) for D-Wave quantum annealers in 2016 \cite{pakin}. This assembler specifically targets quantum annealers like D-Wave's. A comparable technology that targets continuous-variable (CV) quantum computing is the Blackbird language developed by Xanadu since 2018 \cite{2019Quant...3..129K}. We implement parsers for each of these languages in TypeScript with a singular approach. In the cases of Blackbird and QMASM these are the first parser implementations that are web compatible and so bring these languages to a new audience and to new runtimes. This makes the parsing and execution of QMASM, QASM and Blackbird possible in web and mobile environments that don't have access to heavy compile toolchains, enabling adoption and scientific research.

\end{abstract}

\section{Background}

\subsection{Gate Model Quantum Computing}

We can arrive at the specific computational basis vectors typically used for quantum computing in the gate model by considering the Pauli spin operators $\hat{S}$ of a spin-$\frac{1}{2}$ particle such as an electron.

\begin{align}
    S_x = 
    \frac{\hbar}{2} 
    \begin{bmatrix}
        0 & 1 \\
        1 & 0
    \end{bmatrix}\;    &    
    S_z = 
    \frac{\hbar}{2} 
    \begin{bmatrix}
        1 & 0 \\
        0 & -1
    \end{bmatrix} \nonumber
\end{align}

\[
    S_y = -i S_x S_z
\]

\vspace{5mm}

We take our computational basis vectors $\ket{l, m_s} \in \{\ket{0}, \ket{1}\}$ to be defined by the eigenvalue equation $S_z \ket{l, m_s} = \pm \frac{\hbar}{2} \ket{l, m_s}$.

\begin{align}
    \ket{1} = \begin{bmatrix}
        0 \\
        1
    \end{bmatrix}\;\;    &    \;\;\ket{0} = \begin{bmatrix}
        1 \\
        0
    \end{bmatrix} \nonumber
\end{align}

\vspace{5mm}

Together with matrix multiplication, a few other matrices and the CNOT gate, the Pauli matrices form the Clifford algebra. The Clifford algebra plus the so-called $T$ gate is the mathematical backdrop of quantum computing in most cases, though there are alternative universal gatesets \cite{mermin_2016, barenco_bennett_cleve_divincenzo_margolus_shor_sleator_smolin_weinfurter_1995}.

In terms of the basis vectors $\{\ket{0}, \ket{1}\}$ we can write a quantum (qubit) state as $\ket{\psi}$. $\ket{\psi} = \alpha \ket{0} + \beta \ket{1} \; ; \alpha, \beta \in \mathsf{C}$.

Since $\alpha$ and $\beta$ are complex they are normalized by taking their squares. Physically, the squares of these coefficients are the probabilities of observing either basis state via the a measurement in the appropriate basis. $p(\psi = 0) = |\alpha|^2$, $p(\psi = 1) = |\beta|^2$, $|\alpha|^2 + |\beta|^2 = 1$, 

\begin{align}
    M_0 = \begin{bmatrix}
        1 & 0 \\
        0 & 0
    \end{bmatrix}\;\;\;    &    \;\;\;M_1 = \begin{bmatrix}
        0 & 0 \\
        0 & 1
    \end{bmatrix} \nonumber
\end{align}
\vspace{5mm}

It is a simple process to normalize a quantum state. The normalization requirement restricts our state vector to a sphere called the Bloch sphere. This makes it natural to express the state in polar coordinates. $\ket{\psi_{norm}} = 1 / \sqrt{|\alpha|^2 + |\beta|^2} \ket{\psi}$, $|\alpha|^2 = cos^2(\frac{\theta}{2})$, $|\beta|^2 = sin^2(\frac{\theta}{2})$, $\ket{\psi} = cos(\frac{\theta}{2}) \ket{0} + sin(\frac{\theta}{2}) e^{i \phi} \ket{1}$.

The set of matrices that take a state on the Bloch sphere to another state on the Bloch sphere are the same that rotate a vector on a unit sphere. These are the unitary transformations $U$ with the following form and properties. $U$ rotates a state vector by $\gamma$ about $(n_x, n_y, n_z)$. $U^\dagger U = I$, $U^\dagger = (U^*)^T$, $U = e^{i\delta} [cos(\frac{\gamma}{2}) I + i sin(\frac{\gamma}{2})(n_x S_x + n_y S_y + n_z S_z)]$.

\subsection{Continuous Variable Quantum Computing}

The continuous-variable model of quantum information deals with operators that have continuous spectra. Some fundamental operators in quantum physics are not efficient to capture using discrete simulations, like position and momentum. Continuous-variable quantum computing is a unique and valuable computing paradigm since it can simulate physical systems that deal with similarly continuous quantum operators. Rather than dealing in unitary operations, continuous-variable computing constructs a quantum circuit model using continuous gates including quadrature operators, mode operators, state rotations, displacements, squeezing, etc. Rather than a unit of quantum information being represented by a qubit, the basic computational element is a qumode. A qumode has a simple mathematical representation: $\ket{\psi} = \int dx \; \psi(x) \; \ket{x}$.

\vspace{5mm}

Like with a qubit, $\ket{x}$ are eigenstates. In the place of the qubit's discrete coefficients, a qumode has a continuum. A state of quantum information in the CV paradigm is a Gaussian state of the following form, where H is a Bosonic Hamiltonian describing the state in relation to the vacuum state $\ket{0}$. $\ket{\psi} = e^{-i t H} \ket{0}$.

\vspace{5mm}

Like in the qubit system, a CV gate operates on a state. There is a universal set of CV gates as well, which includes the following operators.

\vspace{5mm}

\begin{center}
\begin{tabular}{ c c }
 Displacement & $D_i(\alpha) = e^{\alpha \hat{a}_i^{\dagger} - \alpha^{*} \hat{a}_i}$ \\ 
 Rotation & $R_i(\phi) = e^{i\phi \hat{n}_i}$ \\  
 Squeezing & $S_i(z) = e^{\frac{1}{2}(z^{*} \hat{a}_i^2 - z \hat{a}_i^{\dagger 2})}$ \\
 Beam Splitter & $BS_{ij}(\theta, \phi) = e^{\theta (e^{i\phi}\hat{a}_i^{\dagger} \hat{a}_j - e^{i \phi} \hat{a}_i \hat{a}_j^{\dagger})}$ \\
 Cubic Phase & $V_i(\gamma) = e^{i_{\frac{\gamma}{6}} ]\hat{x}_i^3}$
\end{tabular}
\end{center}

\vspace{5mm}

Unlike qubits in gate model quantum computers, CV states can be measured in different ways. The first type of measurement, a ``homodyne measurement”, is similar to the measuring operator of the qubit and projects a state onto eigenstates. These eigenstates are the eigenstates of the quadrature operator $\hat{x}$: $\hat{x} = \sqrt{ \frac{\hbar}{2}} (\hat{a} +  \hat{a}^{\dagger})$.

\vspace{5mm}

Another type of measurement, a ``heterodyne measurement”, is an uncertain measurement of two operators that do not commute which will yield some probabilistic result for each.

A final type of measurement is a ``photon counting” measurement, which yields the number of photons in a given qumode. This measure projects a state onto the number states. $|\alpha> = e^{-\frac{|\alpha|^2}{2}} \sum_{n=0}^\infty \frac{\alpha^n}{\sqrt{n!}} |n>$.

\subsection{Quantum Annealing}

Quantum annealing is a physical method of minimizing a given function that exploits quantum tunneling to find probable min-terms. D-Wave’s annealers are capable of solving problems of a particular form.

\[
    \mathbf{H}(\hat{\sigma}) = \sum_{i=0}^{N-1}h_i \sigma_i + \sum_{i=0}^{N-2} \sum_{j=i+1}^{N-1} J_{i,j} \sigma_i \sigma_j 
\]

\vspace{5mm}

This equation expresses a Hamiltonian in terms of linear coefficients $h_i$ and quadratic coefficients $J_{i,j}$. In reality, each $\sigma_i$ is a physical qubit, while $h_i$ is specified by the application of an external field and $J_{i,j}$ are inter-qubit coupling strengths.

To perform a minimization of a given problem, an annealer first initializes each qubit to a superposition of 0 and 1 by means of a strong transverse field. The coefficients are configured to specify the parameters of the specific problem to be solved. Then, the transverse filed is lessened over time, to allow each qubit to relax to a stationary state.

The amount of time required for this process is not universally more efficient than a classical annealer solving the same problem. However, it has been proven that a quantum hardware annealer is certainly more efficient than classical analogues in a few cases. The annealing time used by a D-Wave machine can be between 1 $\mu$s and 2000 $\mu$s and can be configured by the user. An important limitation of a D-Wave annealer is that the physical 
topology of the qubit network is a  two-dimensional mesh of 8 qubit bipartile graphs \cite{6802426}. This means that not every qubit may be coupled to every other.

Quadratic pseudo-Boolean functions (QUBOs) of the form in the equation above can be designed such that they are minimized when provided an intended set of relations of inputs and outputs.

\section{Implementations}

The packages share a similar layout. Each provides a user with two interfaces: one for parsing files and one for parsing code strings directly. One might make use of the  packages  together  to  parse  code  targeted  at different backends in a single script by including these imports.

\begin{lstlisting}
import * as qasm from 'qasm-ts';
import * as blackbird from 'blackbird-ts';
import * as qmasm from 'qmasm-ts';
\end{lstlisting}

Each package internally implements a lexer and token parser.  However, the user simply calls either the \textit{parse()} or \textit{parseString()} method to invoke the entire parsing process which includes the steps:

\begin{itemize}
    \item lexing - scanning the provided code (in some cases with a lookahead) and converting it into a list of tokens
    \item parsing - walking the token list and creating an abstract syntax tree representation of the code
\end{itemize}

The tokens are implemented as elements of a TypeScript \textit{enum} enumeration.

\vspace{5mm}
\begin{lstlisting}
enum Token {
    Plus,
    Minus,
    Times,
    Divide,
    ...
}
\end{lstlisting}

These  tokens  are  matched  to  elements  of  the  provided  code  via  a  lookup map.

\vspace{5mm}
\begin{lstlisting}
const lookupMap: object = {
    ',': Token.Comma,
    ':': Token.Colon,
    '"': Token.Quote,
    '(': Token.Lbrac,
    ...
}
\end{lstlisting}

The parser makes use of a class \textit{AstNode} to build the abstract syntax tree. \textit{AstNode} is a parent class for all valid AST elements.  For example, the \textit{Array} class  in  blackbird-ts  is  an  extension  of  a \textit{Parameter} class  which  extends  the \textit{AstNode} class. Once the parser finishes building the AST, it returns a list of nested \textit{AstNodes} to the user.

The \textit{lex()} method of the lexer reads the input file one character at a time and attempts to match string literals to the file contents. In many cases, such as with the string literal for exponentiation in Blackbird (**), a lookahead is required to match the multi-character symbol. The lexer maps symbols in the input file to \textit{Tokens}.

The \textit{parse()} method of each package calls a \textit{parseNode()} method which delegates the parsing of the next \textit{Token} to an appropriate method on the \textit{Parser} class. These vary in complexity from \textit{parseExpression()} which parses a logical or mathematical expression to \textit{bool()} which parses a boolean variable declaration / initialization. Below we include the pseudocode for parseExpression.
\vspace{5mm}
\begin{lstlisting}[language=Java]
/**
 * Parses a logical and mathematical expression.
 * @param tokens - Expression tokens to parse.
 * @return A parsed expression.
 */
parseExpression(tokens: Array<Token>): Expression
    elements = []
    while (tokens.length > 0)
        if (notNestedParam(tokens[0]))
            node = parseNode(tokens)
            if (node != undefined)
                for (let i in node)
                    elements.push(node[i])
            tokens = tokens.slice(1)
        else
            node = parseNode(tokens)
            if (node != undefined)
                for (let i in node)
                    elements.push(node[i])
            while (!matchNext(tokens, Token.Rbrac) && tokens.length > 0)
                tokens = tokens.slice(1)
            tokens = tokens.slice(1)
    return new Expression(elements)
\end{lstlisting}
\vspace{5mm}
The \textit{parseNode()} method maps \textit{Tokens} to objects. These objects include all of the supported gates. The following operators are supported by the blackbird-ts package: Xgate, Zgate, Dgate, Sgate, Rgate, Pgate, Vgate, Kgate, Fouriergate, CXgate, CZgate, CKgate, BSgat, S2gate, Interferometer, GaussianTransform, Gaussian. These state preparations are also supported: Fock, Coherent, Squeezed, Vac, Thermal, DisplacedSqueezed, Catstate. 

A similar set of gates is supported by the qasm-ts package: x, y, z, u1, u2, u3, s, sdg, h, tdg, cx, cy, cz, t, ccx, reset, cu1, ccy, ccz. Qasm-ts also supports custom gates since version 1.1.0.

Qmasm-ts supports the following features of QMASM.

\begin{itemize}
    \item Specifying 2-local Ising Hamiltonian parameters via: qubit weights, coupling strengths
    \item Relating qubits via: chains, anti-chains, equivalences
    \item Relating qubits to classical values via pins
    \item Importable and parameterizable macro system via: \!begin\_macro, include, \!use\_macro
    \item Macro chaining via: \!next
    \item Assertions
    \item Logical and mathematical expressions
    \item For loops
    \item If/else conditionals
    \item Support for the following classical types: Iterator, Range, Int, Float, Bool
    \item Support for the following quantum data types: Qubit, Ancilliary, QubitArray, Register
\end{itemize}

These features are supported by helper functions in the parser class such as \textit{macro, if, assert, etc.} which each help to parse the corresponding feature.

\section{Discussion}

JavaScript (JS) has traditionally been used primarily for implementing re-activity in web frontends, and like Python is not a strongly typed language.  As a result, the language’s implementation does not natively lend itself especially well towards object oriented programming (OOP). JS moves extremely quickly and is arguably the most popular programming language of any kind today \cite{16_OGrady_Name_2023}. With its speed and flexibility come some trade-offs.  For example, JavaScript is arguably the programming language that supports the most conflicting programming paradigms \cite{Delev2018ModernJF}.

However,  the  flexibility  of  JavaScript  yielded  the  expectation  that  it  can be run anywhere. This has made optimizing JavaScript for constrained environments a focus for the JavaScript community.  These efforts have yielded various methods of code minimization, optimization and portability that have become ubiquitous in web programming.  It is also one of the motivations behind the creation of WebAssembly  (WASM) \cite{10.1145/3062341.3062363}.   WebAssembly  is  a  language  that  addresses  the safety, speed and portability of web code. WASM is primarily used as an intermediate compilation target for JavaScript. A future direction for this work might be to produce an optimized WASM distribution of each parser.

Another JavaScript derivative language is Microsoft’s TypeScript. TypeScript was developed with the goal of introducing an optional types system to JavaScript. This makes TypeScript a more appealing language to use than vanilla JavaScript for any application that benefits from typing.  Language definition is definitely one such application.  Microsoft has begun writing experimental languages in TypeScript as well.  For example,  Microsoft published a new language called Bosque that is written exclusively in TypeScript \cite{marron2019regularized}.

What  is  the  connection  between  the  web  and  quantum  computing? Hybrid quatnum-classical algorithms and quantum cloud computing are  the  answer.  In a theoretical hybrid quantum cloud computing scenario, a programmer of an end-user application desires to write an application that can be accessed through a typical mobile or web interface. He/she expects a commercial quantum cloud computing resource to at least be accessible as a service through an understandable web based Application Programming Interface (API). Since QMASM, QASM and BlackBird are programmer’s entry points to using IBM, Xanadu and D-Wave’s quantum systems, bringing these into the world of the web via npm packages is a step towards making hybrid quantum cloud adoption a possibility.

\section*{Acknowledgment}

Thank you for 550 downloads in the last year.

\bibliographystyle{plain}
\bibliography{main}

\end{document}